\documentclass[journal]{IEEEtran}
\usepackage{graphicx}
\usepackage{amsmath}

\begin{document}
\title{Development and Performance Verification\\
of the GANDALF High-Resolution Transient\\
Recorder System}

\author{S.~Bartknecht, H.~Fischer, F.~Herrmann, K.~K\"onigsmann, L.~Lauser, C.~Schill, S.~Schopferer, H.~Wollny
\thanks{Manuscript received June 30, 2010. Revised March 3, 2011.}
\thanks{This work is supported by BMBF, the University of Freiburg and the European Union.}%
\thanks{All authors are with the University of Freiburg, Department of Physics, 79104 Freiburg, Germany.}%
\thanks{Corresponding author: Sebastian Schopferer (telephone: +49-761-203-5877, e-mail: sebastian.schopferer@cern.ch).}%
\thanks{\textcopyright\:2011 IEEE. Personal use of this material is permitted. Permission from IEEE must be obtained for all other uses, in any current or future media, including reprinting/republishing this material for advertising or promotional purposes, creating new collective works, for resale or redistribution to servers or lists, or reuse of any copyrighted component of this work in other works.}%
\thanks{DOI: 10.1109/TNS.2011.2142195}%

}

\maketitle
\thispagestyle{empty}

\begin{abstract}
With present-day detectors in high energy physics one often faces fast analog pulses of a few nanoseconds length which cover large dynamic ranges. In many experiments both amplitude and timing information have to be measured with high accuracy. Additionally, the data rate per readout channel can reach several MHz, which leads to high demands on the separation of pile-up pulses.

For an upgrade of the COMPASS experiment at CERN we have designed the GANDALF transient recorder with a resolution of 12bit@1GS/s and an analog bandwidth of 500\:MHz. Signals are digitized with high precision and processed by fast algorithms to extract pulse arrival times and amplitudes in real-time and to generate trigger signals for the experiment. With up to 16 analog channels, deep memories and a high data rate interface, this 6U-VME64x/VXS module is not only a dead-time free digitization unit but also has huge numerical capabilities provided by the implementation of a Virtex5-SXT FPGA. Fast algorithms implemented in the FPGA may be used to disentangle possible pile-up pulses and determine timing information from sampled pulse shapes with a time resolution better than 50\:ps.
\end{abstract}


\section{Introduction}
\IEEEPARstart{T}{he} COmmon Muon and Proton Apparatus for Structure and Spectroscopy (COMPASS) at the CERN SPS \cite{COMPASS1} is a state-of-the-art two stage magnetic spectrometer \cite{COMPASS-NIM} with a flexible setup to allow for a rich variety of physics programs to be performed with secondary muon or hadron beams. Common to all measurements is the requirement for highest beam intensity and interaction rates with the needs of a high readout speed. Recently a proposal has been submitted \cite{COMPASS2} for studies of Generalized Parton Distributions (GPD), which combine both nucleon electromagnetic form factors and Parton Distribution Functions. Constraining quark GPDs experimentally by measuring exclusive Deeply Virtual Compton Scattering (DVCS) shows great promise for the disentanglement of the nucleon's spin budget. For the upcoming DVCS measurements the existing COMPASS spectrometer will be extended by a new 2.5\:m long liquid hydrogen target, which will be surrounded by a new recoil proton detector based on scintillating counters. The high luminosity of about $10^{32}\:\text{cm}^{-2}\text{s}^{-1}$ and the background induced by the wide beam halo will yield rates of the order of several MHz in the recoil detector counters. This imposes great demands on the digitization units and on a hardware trigger based on the recoiled particle. For this purpose we have developed within the GANDALF framework \cite{GANDALF-NIM} a modular high speed and high resolution transient recorder system featuring digital pulse processing in real-time.

\begin{figure}[!t]
\centering
\includegraphics[width=3.47in]{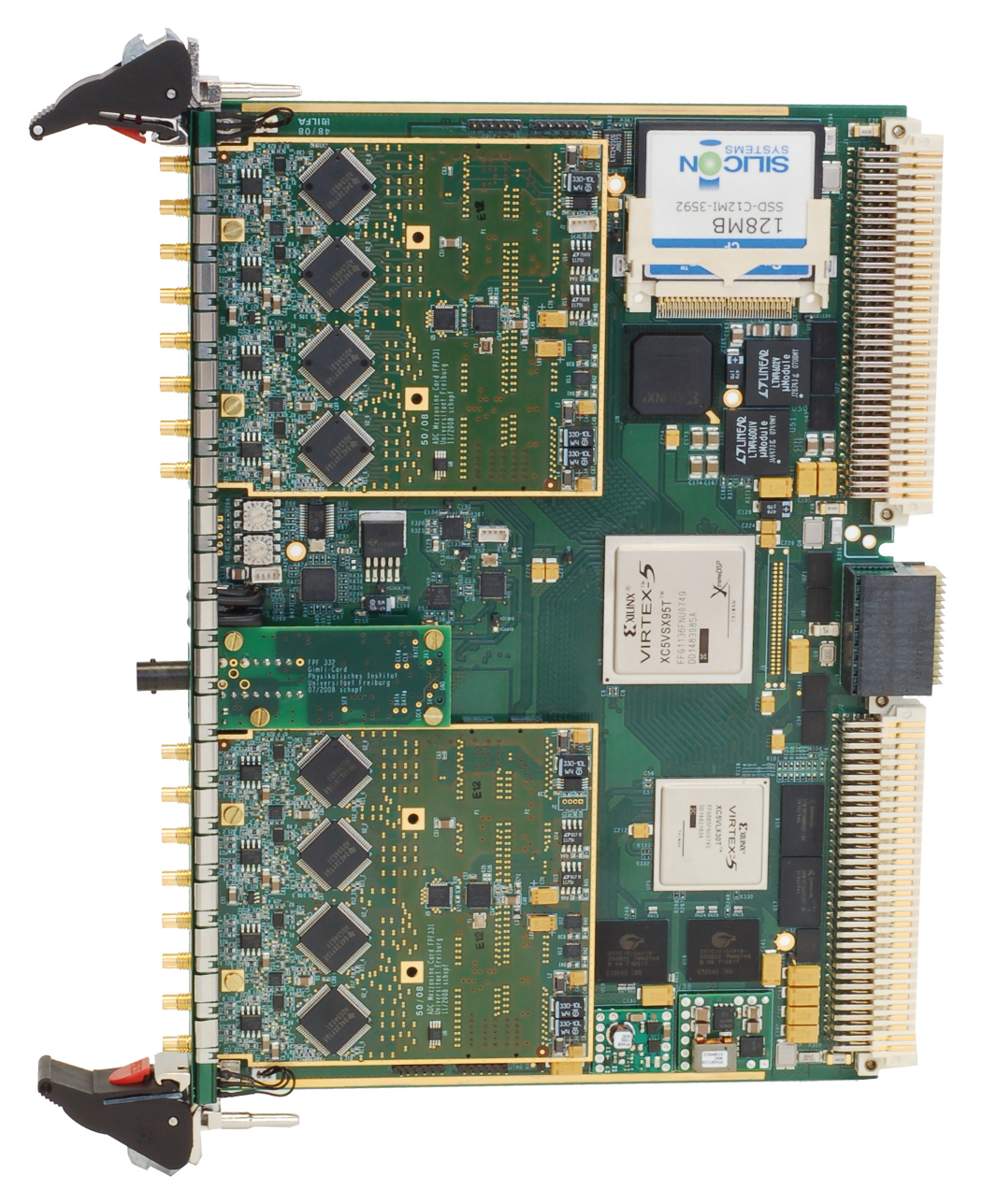}
\caption{Picture of the GANDALF carrier board equipped with two ADC mezzanine cards. The center mezzanine card hosts an optical receiver for the COMPASS trigger and clock distribution system.}
\label{fig_gandalf}
\end{figure}

\section{The GANDALF Framework}

GANDALF (Fig. \ref{fig_gandalf}) is a 6U-VME64x/VXS \cite{VITA} carrier board which can host two custom mezzanine cards. It has been designed to cope with a variety of readout tasks in high energy and nuclear physics experiments. The exchangeable mezzanine cards allow an employment of the system in very different applications such as analog-to-digital or time-to-digital conversions, coincidence matrix formation, fast pattern recognition or fast trigger generation. Currently two types of mezzanine cards are available: ADC cards and LVDS input cards. Another model with optical interfaces is foreseen to receive data from remote detector frontend modules. 

\begin{figure}[!t]
\centering
\includegraphics[width=3.47in]{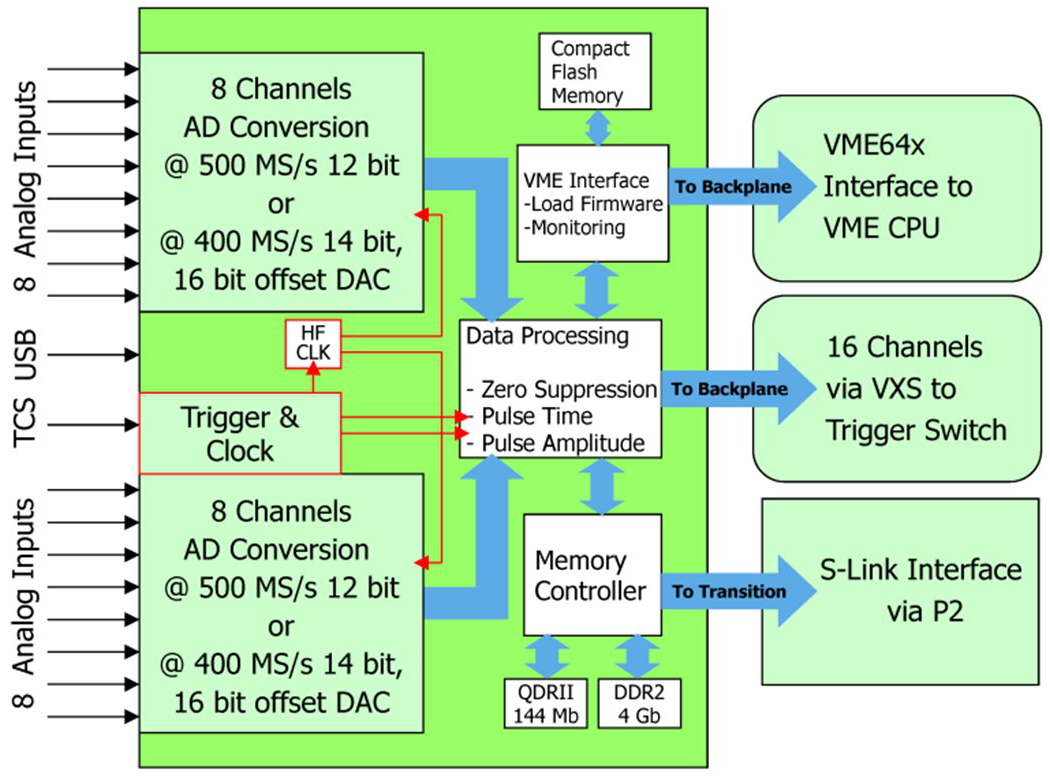}
\caption{Block diagram of GANDALF as a transient recorder.}
\label{fig_gandalf_scem}
\end{figure}

When GANDALF is used as a transient recorder, the carrier board is equipped with two ADC mezzanine cards. A schematic overview is provided in Fig. \ref{fig_gandalf_scem}. The heart of the board is a Xilinx VIRTEX5-SXT FPGA which is connected to each mezzanine card by several single ended and 120 differential signal interconnections. The data processing FPGA can perform complex calculations on data which have been acquired on the mezzanine cards to extract time and amplitude information of the sampled pulses.

Fast and deep memory extensions of 144-Mbit QDRII+ and 4-Gbit DDR2 RAM are connected to a second Virtex5 FPGA. Both FPGAs are linked to each other by eight bidirectional high-speed Aurora lanes with a total bandwidth of 25\:Gbit/s per direction.

Connected to the VXS backplane GANDALF has 16 high-speed lanes for data transfer to a central VXS module, where the lanes of up to 18 GANDALF modules merge. This connection can be used for continuous transmission of the amplitudes and the time stamps from sampled signals to the VXS trigger processor, which then forms an input to the experiment-wide first-level trigger based on the energy loss and the time-of-flight in the recoil detector.

A dead-time free data output can either be realized by dedicated backplane link cards connected to each GANDALF P2-connector, i.e. following the 160 MByte/s SLink \cite{SLINK} or Ethernet protocol, or by the VME64x bus in block read mode \cite{lauser} or by USB2.0 from the front panel. Depending on requirements the data output may contain the results of the digital pulse processing only or also the full sample list.

\section{Analog-to-Digital Converter}

Two models of analog-to-digital converters (ADC) can be used with the GANDALF board, depending on the desired resolution. With the Texas Instruments models ADS5463 (12bit@500MS/s) and ADS5474 (14bit@400MS/s) we chose two of the fastest pipelined high resolution ADC chips that are currently available. Their low latency of only 3.5 clock cycles gives valuable time for the signal processing and the following trigger generation with its tight timing constraints defined by existent readout electronics.

\begin{figure}[!t]
\centering
\includegraphics[width=3.47in]{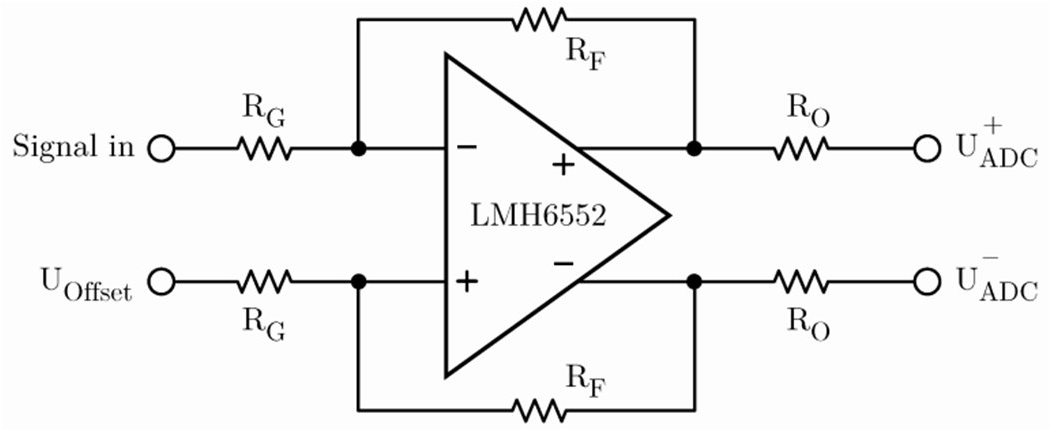}
\caption{Schematic of the DC-coupled analog input circuit. For each channel $\text{U}_\text{Offset}$ can be set by 16-bit DACs.}
\label{fig_analog}
\end{figure}

The DC-coupled analog input circuit (Fig. \ref{fig_analog}) uses the differential amplifier LMH6552 from National Semiconductor and has a bandwidth of 500\:MHz. It adapts the incoming single ended signal, e.g. from a photomultiplier tube (PMT), to the dynamic range of the ADC while the baseline of each channel can be adjusted individually by 16-bit digital-to-analog converters (DAC). Depending on the DAC settings, the input stage accepts unipolar or bipolar pulses.
Two adjacent channels can be interleaved to achieve an effective sampling rate of 1\:GS/s (800\:MS/s with the ADS5474) at the cost of the number of channels per mezzanine card. In this time-interleaved mode the second ADC receives a sampling clock which is phase-shifted by 180 degree and the input signal is passively split to both channels. Thus the signal is sampled alternately by two ADCs.

\begin{figure}[!t]
\centering
\includegraphics[width=3.47in]{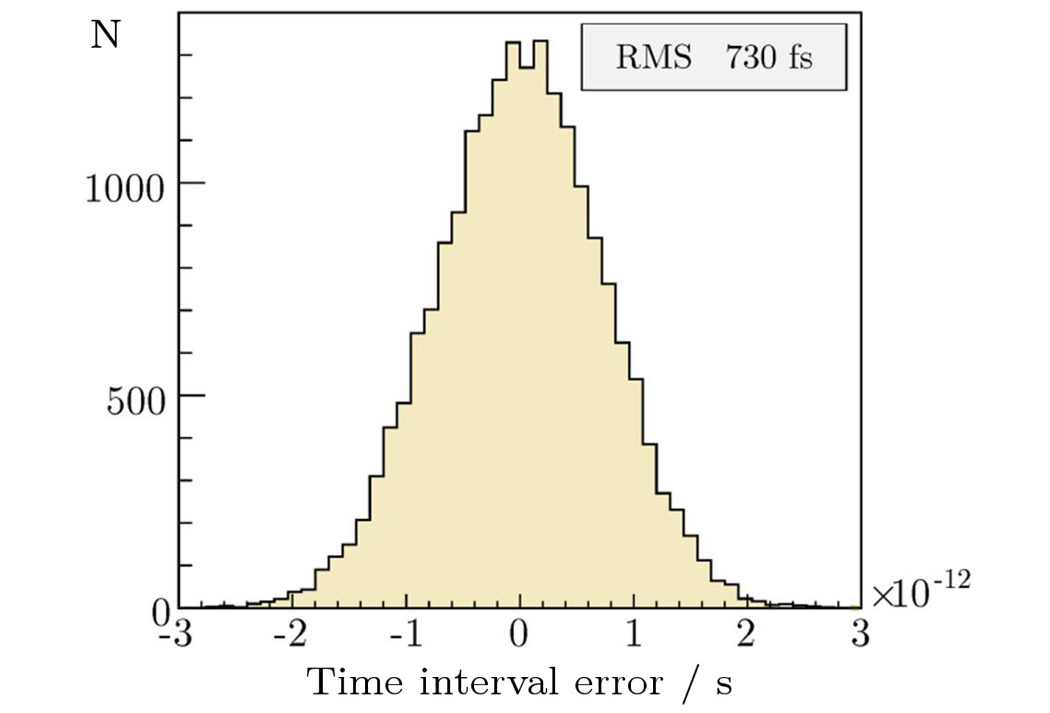}
\caption{Time interval error of the sampling clock. The measurement was performed with a Tektronix TDS6154C using the TDSJIT3 software.}
\label{fig_jitter}
\end{figure}

On each ADC mezzanine card the high frequency sampling clock is generated by a digital clock synthesizer chip SI5326 from Silicon Labs, which comprises an integrated PLL consisting of an oscillator, a digital phase detector and a programmable loop filter. The experiment-wide 155.52-MHz clock, distributed by the COMPASS trigger and clock distribution system (TCS), is used as reference. Particular attention has been paid to the design of the clock filter networks and the board layout to reach a time interval error smaller than 730\:fs (Fig. \ref{fig_jitter}) \cite{schopf}, which is essential for high bandwidth sampling applications.

\begin{figure}[!t]
\centering
\includegraphics[width=3.47in]{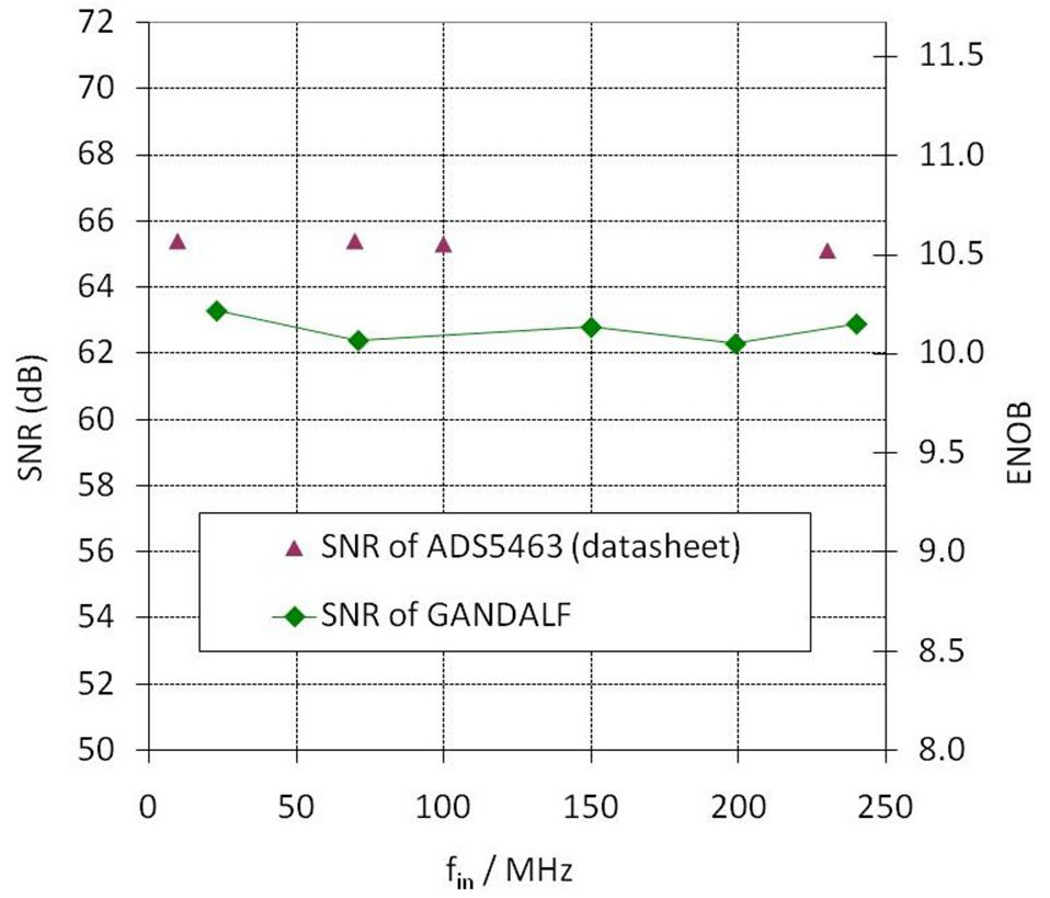}
\caption{Signal-to-noise ratio (full-scale) and effective resolution of the 12-bit GANDALF digitization unit using the ADS5463 ADC at 500\:MHz. Values from the ADS5463 datasheet are given for comparison for selected analog input frequencies.}
\label{fig_snr}
\end{figure}

We determined the signal-to-noise ratio (SNR) of the ADC system in a test setup consisting of a high precision function generator (Tektronix AFG3252) and a selection of narrow band pass filters. The filters were connected directly to the analog input of the GANDALF module to suppress the harmonics of the signal source. Sine waveforms of different frequencies were sampled and from the fast Fourier transform the SNR was calculated. The result of these measurements for the 12-bit version (ADS5463) is shown in Fig. \ref{fig_snr} as a function of the frequency of the input analog signal and is expressed in dB as well as ENOB (effective number of bits). We achieved an effective resolution of above 10.1\:ENOB (ADS5463) and 10.6\:ENOB (ADS5474) respectively over an input frequency range up to 240\:MHz.

\section{Digital Pulse Processing}

The sampled detector signals are processed by DSP algorithms inside the VIRTEX5-SXT FPGA utilizing the high compute power of its 640 DSP48E Slices. Quantities of interest like pulse arrival time, pulse height and integrated charge are extracted and can be used for real-time calculation of derived quantities such as time-of-flight and energy loss.

\subsection{Pulse Time Determination}

\begin{figure}[!t]
\centering
\includegraphics[width=3.47in]{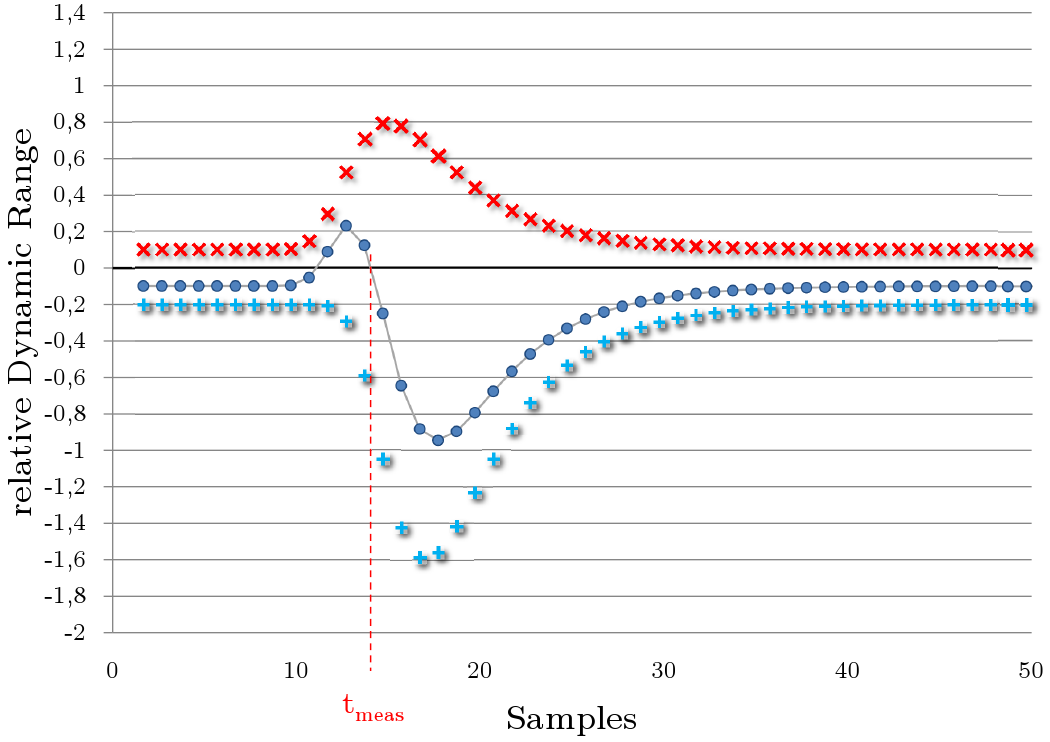}
\caption{Illustration of the digital constant fraction (dCF) method. The original samples ($\times$) and the delayed and inverted samples ($+$) add up to the dCF function ($\bullet$).}
\label{fig_cfd}
\end{figure}

To extract the time information from the detector signals a digital constant fraction discrimination (dCFD) algorithm was chosen. Inside the FPGA the digitized samples are delayed, multiplied by a fraction factor and added to the original samples (Fig. \ref{fig_cfd}). The zero-crossing of the resulting curve is determined by linear interpolation and forms the time stamp of the pulse. Extended simulations \cite{bartkn} helped to determine optimal parameters for the dCFD algorithm also in case of pile-up pulses.

\subsection{Performance Verification}

\begin{figure}[!t]
\centering
\includegraphics[width=3.47in]{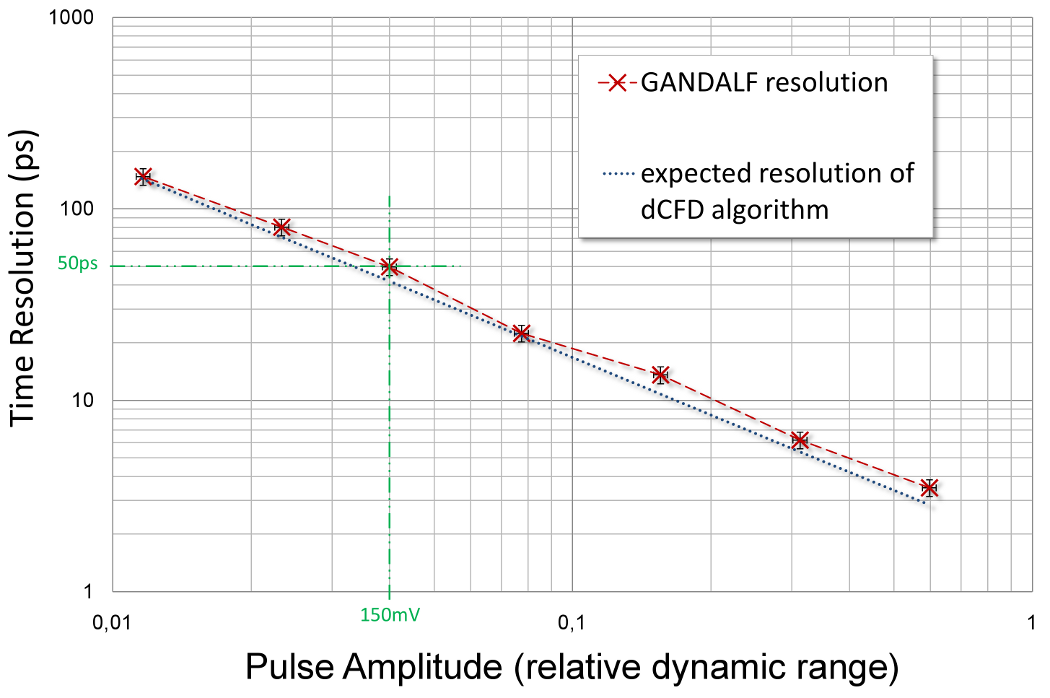}
\caption{GANDALF timing resolution measured with pulses of 2\:ns risetime generated by an AFG3252 arbitrary function generator. The dynamic range in this case is 0\:V to -4\:V. The expected resolution of the dCFD algorithm is determined by simulation.}
\label{fig_tim1}
\end{figure}

First measurements of the timing resolution of the GANDALF digitizer were performed by using a Tektronix Arbitrary Function Generator (AFG3252) to simulate realistic detector pulses. The PMT signals are described by a moyal distribution
\[
	f(t) = A \cdot exp \left(-\frac{1}{2} \left(\frac{t-t_0}{k \cdot t_r} + exp\left(-\frac{t-t_0}{k \cdot t_r}\right) -1\right)\right),
\]
with maximum amplitude $A$ at $t=t_0$. For $k=0.69$, $t_r$ is the 10\%-90\% rise time. Two copies of the signal with constant but arbitrary delay are sampled and the constant fraction timing is performed by the DSP FPGA. The differences between the resulting time stamps show a distribution, whose width is then used to calculate the timing resolution. The dCFD resolution depends on the pulse amplitude. Therefore the measurements were done with a signal amplitude variation over the dynamic range of the input stage. In this test setup pulses with amplitudes from -50\:mV to -4\:V were measured with 1\:GS/s (Fig. \ref{fig_tim1}). One can see that with the GANDALF transient recorder a timing resolution of better than 50\:ps can be reached for signal amplitudes as small as 4\% of the relative dynamic range.

\begin{figure}[!t]
\centering
\includegraphics[width=3.47in]{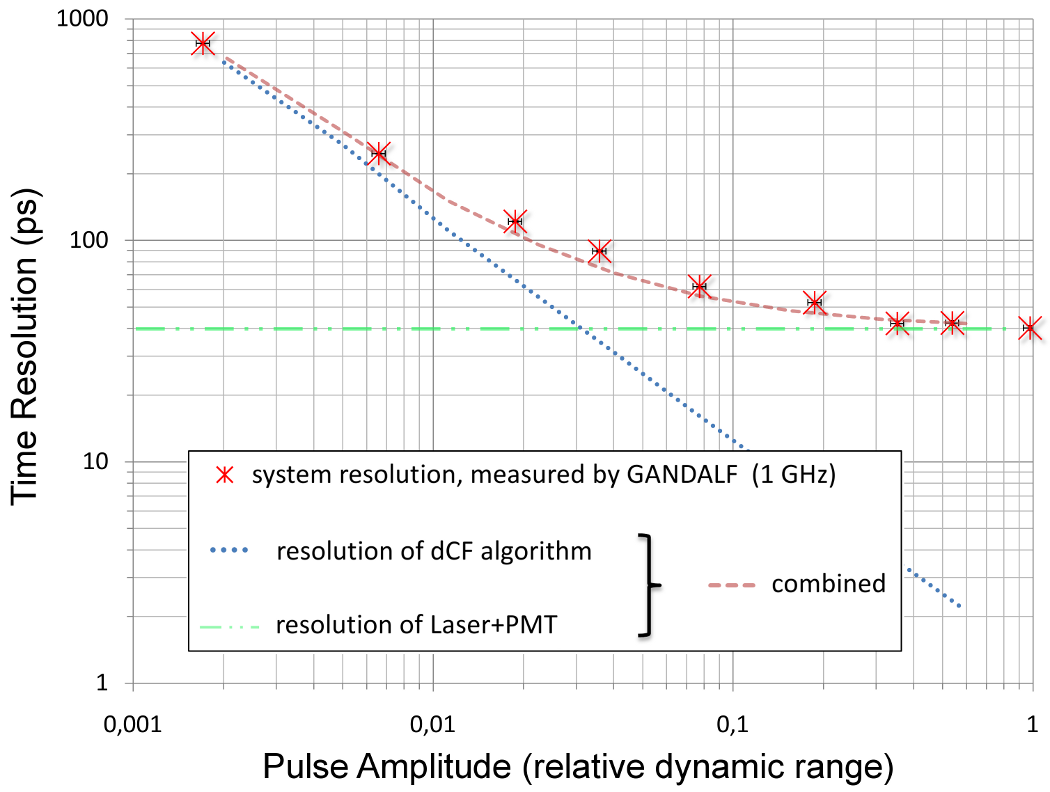}
\caption{Timing resolution of PMT signals generated from a Picosecond Injection Laser and measured with GANDALF.}
\label{fig_tim2}
\end{figure}

To confirm these results in a real environment a measurement setup with a Hamamatsu PMT (R1450) as a signal source was installed. In this configuration a PiLas EIG1000D laser pulser with PiL040 Optical Head (from Advanced Laser Diode Systems) is used as light source. The laser emits very short optical pulses with pulse widths below 45\:ps (FWHM). The pulser features an additional TTL trigger output, which is used as a time reference. The jitter between the trigger and the optical output is typically below 3\:ps. In Fig. \ref{fig_tim2} the timing resolution of the laser and PMT system measured by GANDALF in 1\:GS/s mode is shown. The resolution of the system as a whole (green continuous line) is composed of the resolution of the GANDALF timing determination (blue dashed line) and the resolution of the laser and PMT system (red dash-dotted line). The latter is quantified in an independent measurement to be 39\:ps.

\begin{figure}[!t]
\centering
\includegraphics[width=3.47in]{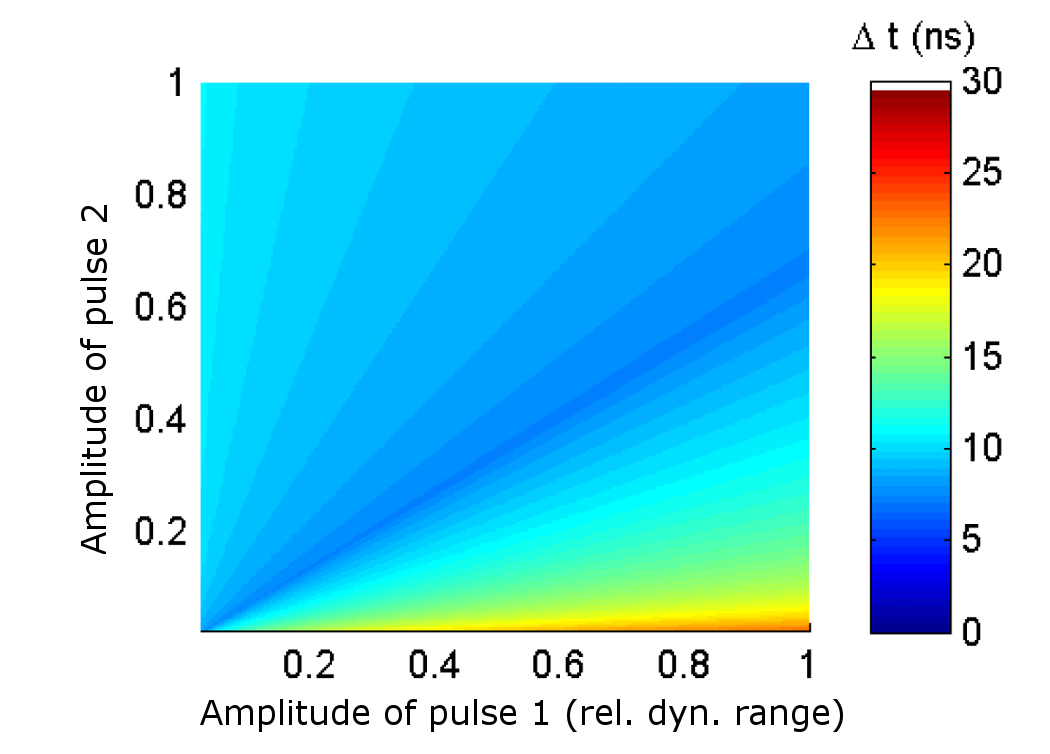}
\caption{Pulse separation ability of the dCFD algorithm depending on the amplitudes of the first (x-axis) and second (y-axis) pulse. The color code denotes the minimum delay between the pulses which is required for separation.}
\label{fig_doublepulses}
\end{figure}

The ability of the dCFD algorithm to separate pile-up pulses depends on the ratio of the pulse amplitudes. Fig. \ref{fig_doublepulses} shows the minimum delay $\Delta t$ between two pulses which is needed to separate them. The plot is obtained by simulation of two consecutive pulses with risetime $t_r=3\:ns$ for all combinations of amplitudes of the first and second pulse. The simulation result is verified by lab measurements using the AFG3252 to generate pile-up pulses with selected amplitude ratios.

\section{Conclusion}

A low cost VME64x system aimed at digitizing and processing detector signals has been designed and implemented to our full satisfaction. The design is modular, consisting of a carrier board on which two mezzanine boards with either analog or digital inputs can be plugged. The ADC mezzanine cards have been characterized and show excellent performance over a wide input frequency range. An optional high-speed serial VXS backplane offers inter-module communication for sophisticated trigger processing covering up to 288 detector channels.

The GANDALF transient recorder has been installed at the COMPASS experiment during a two-week DVCS pilot run in September 2009. Extensive data have been recorded in order to verify the performance of the hardware and the signal processing algorithms.

\section{Outlook}

Recently an additional type of mezzanine card with 64 digital inputs has been designed, which accepts LVDS and LVPECL signals over a VHDCI connector. Using this digital mezzanine cards a 64-channel mean-timer and subsequent trigger matrix was implemented \cite{bieling} in the GANDALF module and is in action at COMPASS since April 2010.

In a forthcoming paper we will describe the realization of GANDALF as a 128-channel time-to-digital converter module with 100\:ps digitization units, comparable to the F1-TDC chip \cite{fischer}. The TDC design is implemented inside the main FPGA which can host 128 channels of 500-MHz scalers at the same time.

\section*{Acknowledgment}
The authors gratefully acknowledge the discussions with their colleagues from the COMPASS collaboration and the support of their local workshops.



\end{document}